\documentclass[a4paper, 10pt, conference]{ieeeconf}      

\IEEEoverridecommandlockouts                              
\usepackage{amsmath,amsfonts,amssymb}
\usepackage{mathtools, textcomp}
\usepackage{mathrsfs}
\usepackage{graphicx}
\usepackage[dvips]{epsfig}
\usepackage{epstopdf}
\usepackage{color}
\usepackage{subfig}
\usepackage[compress,sort,noadjust]{cite}
\usepackage{cite}

\title{Chatter Avoidance in Delayed Feedback Attitude Control with MRP Shadow Set Switching 
}

\author{Ehsan Samiei and Eric A. Butcher
\thanks{E. Samiei is with the Department of Mechanical and
Aerospace Engineering, New Mexico State University, Las Cruces,
NM 88003, USA 
        {\tt\small esamiei at nmsu.edu}}%
\thanks{E. A.Butcher is the Faculty member of Aerospace and Mechanical Engineering,
University of Arizona, Tucson, AZ 85721
        {\tt\small ebutcher@email.arizona.edu}}%
}

\begin{document}

\maketitle
\thispagestyle{empty}
\pagestyle{empty}

\begin{abstract}
The chattering response at the MRP shadow set switching point for the controlled attitude dynamics of a rigid tumbling spacecraft using delayed state feedback control with MRPs is investigated, where the time delay is assumed to be in the measurement of the state. In addition, a strategy to reduce or completely avoid the chattering phenomena using a hysteretic boundary layer switching rule is employed. Simulations are performed to demonstrate the chattering phenomenon and the advantages of the modified MRP shadow set switching rule.  
\end{abstract}

\section{Introduction}
Attitude representation depends on the choice of attitude parameters used to represent the orientation of a rigid body relative to an inertial frame see, e.g. \cite{Marandi,Shuster,Schaub, Schaub3}~. Attitude parameters  can be either the rotation matrix, the principal angle and principal axis, Euler angles (EA), Euler parameters  or quaternions (EPs), classical Rodriguez parameters (CRPs) or modified Rodriguez parameters (MRPs), among others~\cite{Schaub}$\:$.  The principal rotation vector is a  basis  for many attitude representations, but it has the disadvantage of  the mathematical singularity  for zero rotation.  EAs are easy to visualize, but the reference frame is never  more than 90 degrees rotation from a  singularity. EPs, on the other hand, have considerable benefits including the facts that they are nonsingular and their corresponding kinematic equation is linear, and hence are widely used in  spacecraft attitude studies. However, they are quite hard to  visualize. CRPs, also known as Gibbs vector, reduce the EPs to a minimal three-parameter set. Based on their definition, they are much better suited for large spacecraft rotations than EAs. A further improved attitude representation,  known as MRPs, moves the singularities to $\mathrm{360^o}$ rotation instead of $\mathrm{180^o}$ as in the case of  CRPs.

On the other hand, in recent decades, the problem of delayed feedback control  has been subject to intensive research because of the wide range of applications in mechanical systems such as spacecraft attitude maneuvers \cite{Schaub1,Sharma04}~, underwater vehicles \cite{Sanyal1}~,  cooperative robot manipulators \cite{karpinska}~, etc. In addition, in some practical applications, there is a time delay within the control system due to the delay in the communication channel or the actuator delay. Few studies have focused on delayed feedback control of attitude dynamics \cite{Chunodkar, Samiei2012, SamieiN, Ailon, Kim, Morad1, Samiei2015attitude, Morad, SamieiIFAC}. A velocity free output-based controller for attitude regulation of a rigid spacecraft considering the effects of a known time delay in the system is presented in \cite{Ailon}~. Sufficient conditions for attitude stabilization of the spacecraft were also obtained. In \cite{Chunodkar}~, a linear state feedback controller with unknown time delay and known upper bound in the feedback path using a frequency domain approach is designed. A complete type Lyapunov-Krasovskii functional was constructed to ensure the stability robustness of linear controller. An estimate of the region of attraction was then obtained and the exponential stability of the system in this region guaranteed. Kim and Crassidis \cite{Kim} implemented a nonlinear robust controller followed by an optimal design algorithm on a spacecraft attitude dynamics without any angular velocity measurements in the presence of the constant time delay in the control signal. The closed loop system is shown to be stable for a norm bounded nonlinear uncertainty. The delayed feedback stabilization of rigid spacecraft
attitude dynamics in the presence of stochastic input torques
and an unknown time-varying delay in the measurement is also addressed in \cite{Samiei2012}. By employing a linear state feedback controller via a Lyapunov-
Krasovskii functional, a general delay-dependent mean-square
stability condition was characterized for the closed-loop parameterized
system in terms of a linear matrix inequality (LMI). A quadratic cost function was  applied to the derived LMI to achieve a suboptimal control performance for the parameterized system. An estimate of the region of attraction of the controlled system is also obtained, inside which the asymptotic stability of the parameterized system is guaranteed in the mean square sense.  

MRPs provide a minimal attitude representation; however they suffer from nonuniqueness  as well as a singularity after one
complete revolution. A unique singularity free representation of attitude can be obtained if switching to the shadow set is implemented. However, in most of the works in the delayed feedback control design where MRPs are employed see, e.g. \cite{Chunodkar,Samiei2012}~, switching has not been considered since it is assumed that the magnitude of the MRP set is always less than one. However, this is a restrictive assumption for the purpose of the control design and needs to be addressed properly. The same issues can emerge when the MRPs are used for the attitude filtering \cite{Keefe}. In this paper, we investigate the chattering phenomenon for the controlled attitude dynamics of a spacecraft in the presence of a constant known time delay in the measurement when the spacecraft is tumbling through the MRP switching point. In addition, the chattering avoidance is also addressed to eliminate this phenomena in the state variables of the system. We assume that there is a constant known time delay in the measurement which causes the time delay to appear in the feedback loop.  Here, the time delay is considered to exist in the sensors, while in ~\cite{Morad}$\:$, the time delay appears in the actuators. By employing a linear state feedback controller to the closed-loop system and switching between the standard and alternate (shadow) MRP sets to describe a certain attitude, asymptotic stability of the parameterized system can be guaranteed for a wide range of attitude maneuvers. The main objective of this paper is to show that there is a significant drawback to this approach when the rigid body is tumbling and switching to the shadow set is employed. The delayed feedback controller can cause chattering phenomenon for the attitude representation, and the asymptotic stability of the system is no longer guaranteed.  To reliably avoid the chattering phenomena, boundary layer solutions are employed to address the asymptotic stability of the system in the finite time. A set of simulations are performed to show that the chattering phenomenon on the delayed feedback control of tumbling spacecraft can happen and the chattering effect can be eliminated by  using the boundary layer method on the response of the closed loop system. 

This paper is organized as follows: In section II, we represent an attitude dynamics model using MRPs and then introduce two sets of MRPs which describe the same orientation. Section III implements  a linear delayed state feedback control law to the parameterized system and then the closed-loop system is obtained in terms of new state variables. In section IV, we investigate the response of the delayed feedback controlled system when the spacecraft is near tumbling situation for two different strategies where either the current or the delayed values of the attitude parameters are available for the measurement and we will show that the chattering phenomena occurs in the response of the controlled system.  To eliminate the chattering phenomenon, in Section V, we implement the boundary layer method to the control design in order to asymptotically stabilize the closed-loop parameterized system. Section VI concludes the paper.       

\section{System Modeling}
\begin{figure}
    \includegraphics[width=0.5\textwidth]{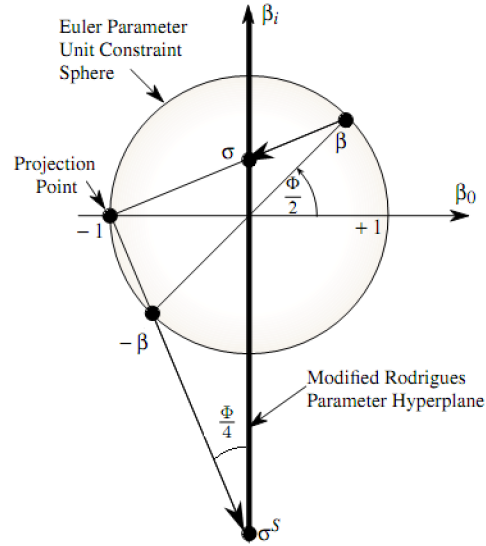}
  \caption{stereographical projection of Euler parameters to MRPs \cite{Schaub}}
  \vspace{-6pt}
\label{stereo}
\end{figure}

In this paper, the spacecraft is modeled as a rigid body expressed in the body-fixed coordinate frame $\mathcal B$. To describe the orientation of the spacecraft relative to the inertial frame $\mathcal N$, MRPs are employed as the set of three-dimensional attitude parameterization. The MRPs, denoted by $\sigma(t)\in\mathcal R^3$, can provide a minimal representation; however they suffer from nonuniqueness as well as a singularity after one complete revolution.  A unique non-singular minimal set of attitude parameters can be obtained if switching to the shadow set when $\left\|\sigma(t)\right\|= 1$ are employed such that $\left\|\sigma(t)\right\| \le 1$ \cite{Schaub}, where $\left\|.\right\|$  represents the Euclidean norm of a vector. More details about different set of attitude coordinates can be found in, see e.g \cite{Marandi,Shuster,Schaub, Schaub3}. The MRP attitude parameterization can be introduced in terms of a nonlinear function of Euler’s principal rotation elements as
\begin{eqnarray}
\sigma=\tan\frac\Phi4\:\hat e,
\end{eqnarray}
where $\hat e$  is a unit eigenvector along the principle axis of rotation and $\Phi$ is the principle rotation angle. As is mentioned in the above, the MRPs are non-unique. Therefore there are two different MRP sets for each attitude. This is due to the fact that principle rotation elements are non-unique, i.e., two different sets of principle rotation elements $(\hat e,\Phi)$ and $(\hat e,2\pi-\Phi)$ can be defined for each attitude. Thus, the MRP shadow set can be obtained using the alternate principle rotation set $(\hat e,2\pi-\Phi)$ as
\begin{eqnarray}
\sigma^s(t)=-\frac{1}{\left\|\sigma\right\|^2}\sigma(t),
\end{eqnarray}
which represents the same orientation as $\sigma(t)$. The original and shadow MRP sets can be also visualized by considering the stereographical projection of Euler parameters to MRPs \cite{Schaub} as illustrated in Fig.(\ref{stereo}), where $\beta_0$ is the scalar part and $\beta=(\beta_1,\beta_2,\beta_3)^T$ is the vector part of the quaternions. The
quaternion constraint $\sum\limits_{i=1}^{3}\beta_i^2=1$ always holds. As is shown in Fig.(\ref{stereo}) when one set of MRPs leaves the unit sphere $\left\|\sigma(t)\right\|=1$ the other set enters the sphere. Thus by performing the switching between the sets we always constrain the MRP set to remain inside the unit sphere while representing the same orientation for the spacecraft.

Considering there are three actuators acting along orthogonal axes in frame $\mathcal B$, the attitude dynamics of a rigid spacecraft can be obtained as
\begin{eqnarray}\label{attitude}
\nonumber \dot {\sigma}(t)=&&\frac 14 B({\sigma(t)}){\omega(t)}\\
\dot {\omega}(t)=&&- J^{-1}{{\omega}^\times}(t)  J{\omega(t)}+ J^{-1} u(t),
\end{eqnarray}
where $\omega(t)\in\mathcal R^3$ is the angular velocity, $ u(t)\in\mathcal R^3$ is the control torque vector,  $J$ is the inertia matrix, and $(\cdot)^\times$ is defined as 
\begin{eqnarray}
\omega^\times=\left(\begin{array}{ccc}0&-\omega_3&\omega_2\\\omega_3&0&-\omega_1\\-\omega_2&\omega_1&0\end{array}\right),
\end{eqnarray}
The first part of  Eq.(\ref{attitude}) is the kinematic differential equation which is analogous to Poisson's equation $\dot C(t)=-\omega(t)^\times C(t)$ where $C(t) \in SO(3)$ is the direction cosine matrix that rotates vectors from $\mathcal N$ to $\mathcal B$, while the second part represents the rotational dynamics (Euler's equation). It should be noted that in Eq.(\ref{attitude}), it is assumed that the individual dynamics of the sensors and actuators are not considered. The nonlinear function $B(\sigma)$ in Eq(\ref{attitude}) can be expressed as
\begin{eqnarray}
B(\sigma)=\left[\left(1-\sigma^T\sigma\right)I_3+2\sigma^\times+2\sigma\sigma^T\right],
\end{eqnarray}
where $I_3$ is the $3\times3$ identity matrix. 
Defining the new state variable $x(t)=[\sigma^T(t),\frac{1}{4}\omega^T(t)]\in \mathbb R^6$, Eq.(\ref{attitude}) can be rewritten in the following form:
\begin{eqnarray}\label{ss}
\dot x(t)= Ax(t)+Bu(t)+f(x(t)),
\end{eqnarray}
where
\begin{eqnarray}\label{constantp}
\nonumber &A&= \left[ {\begin{array}{*{20}{c}}
   {{0_{3 \times 3}}} & {{I_{3 \times 3}}}  \\
   {{0_{3 \times 3}}} & {{0_{3 \times 3}}}  \\
\end{array}} \right]
,~B= \left[ {\begin{array}{*{20}{c}}
   {{0_{3 \times 3}}}  \\
   {\frac{1}{4}{J^{ - 1}}_{3 \times 3}}  \\
\end{array}} \right]\\
&f(x(t))&=\left[ {\begin{array}{*{20}{c}}
   {f_1}  \\
   { f_2}  \\
\end{array}} \right]= \left[ {\begin{array}{*{20}{c}}
   {[ B(x_1(t))-{I_3}]x_2(t)}  \\
   { - 4{J^{ - 1}}\tilde x_2(t) Jx_2(t) }  \\
\end{array}} \right]\in {\mathcal R^6},
\end{eqnarray}
 $x_1(t)$ and $x_2(t)$ are components of state variable $x(t)$, $f(x(t)): \mathcal D\rightarrow R^n$ is a piecewise continuous function, and $\mathcal D\subset R^n$ is a domain of the system.

Eq.(\ref{attitude}) can be linearized about $\sigma=0$ as 
\begin{eqnarray}\label{linearized}
&&\dot \sigma=\frac 14 \omega(t)\\
\nonumber &&\dot {\omega}(t)=J^{-1} u(t)
\end{eqnarray}
Note that the MRP's linearize as angles over four, i.e. $|\vec\sigma|\approx\frac{\Phi}{4}$, so that there is a wide range of attitude maneuvers for which the linearized approximation is valid \cite{Schaub}. This linearizion can help us to design a suitable linear delayed feedback controller for the attitude dynamics which will be discussed in the next following section.

\section{Delayed State Feedback Controller}

Let us consider a linear delayed state feedback controller as
\begin{eqnarray}\label{cs}
u(t)=Kx(t-\tau),
\end{eqnarray}
where $\tau$ represents the constant time-delay, $K=[4JK_1~~JK_2]\in {R^{3 \times 6}}$ and $ K_{1},K_{2} \in {\mathcal R^{3\times 3}}$ are constant controller gain matrices. It is shown in \cite{Samiei2012} that the above state feedback controller can stabilize the parametrized model in Eq. (\ref{ss}) such that all the angular velocities and attitude parameters go to zero as $t \rightarrow \infty$ in some region of the domain $\mathcal D$ that contains the origin in the presence of a time-varying delay in the feedback path, i.e.,
$\lim_{t\rightarrow\infty} \left\|\ x(t)\right\|=0$. Therefore, substituting Eq.(\ref{cs}) into Eq(\ref{ss}), the closed loop system can be obtained in the form of a nonlinear delay-differential equation (DDE) as
\begin{eqnarray}\label{nonc}
&&\nonumber \dot x(t) = Ax(t) + BKx(t - \tau )+f(x(t)),\,\,\,\,\,t \ge 0,\\
&&x(\theta)=\phi(\theta),\,\,\,\,\,\,\,-\tau(t) \le \theta \le 0,
\end{eqnarray}
where $\phi(\theta)$ is the initial function and the infinite dimensional state defined by $x_t(\theta)=x(t+\theta), -\tau \le \theta \le 0$ resides in the Banach space $\mathcal C([-\tau,0],\mathcal R^6)$, and the matrix $BK$ is defined as
\begin{eqnarray}\label{19oct11-0140pm}
\nonumber~~BK= \left[ {\begin{array}{*{20}{c}}
   {{0_{3 \times 3}}} & {{0_{3 \times 3}}}  \\
   {{K_{1}}} & {{K_{2}}}  \\
\end{array}} \right] \in {R^{6 \times 6}}.
\end{eqnarray}

A suitable choice of the control gain matrix $K$ can be obtained by solving a LMI feasibility problem through a Lyapunov-Krasovskii functional candidate (see.e.g., \cite{SamieiL}) as presented in \cite{Samiei2012}$~$. However, switching between the original and shadow MRP sets has not been employed in this study since the assumption $\left\|\sigma\right\| <1,~ \forall t \ge 0$ has been made.

\section{Chattering Behavior at MRP Switching Point}
In this section, we explore the controlled response of Eq.(\ref{nonc}) when the spacecraft is tumbling through the MRP switching point $\left\|\sigma(t)\right\|=1$. Consider a rigid spacecraft with inertia matrix $J=diag(140,100,80)$. It initially starts at $\Phi=171.6913^\circ$ and $\hat e=[1,0,0]^T$ with the initial angular velocity $\omega=[0.46,0,0]^T (\frac{rad}{sec})$. the spacecraft has a large initial MRP and the angular velocity that cause the spacecraft tumble through $\Phi=180^\circ$ about the first body axis. Initial conditions in terms of MRPs can be obtained as $\sigma=[0.93,0,0]^T$.  Let us assume that there is no time delay in the measurement and the spacecraft can be stabilized by the linear controller Eq(\ref{cs}) without considering the time delay, i.e., $\tau=0$ and $u(t)=Kx(t)$. To obtain the control gain matrix $K$ for this scenario, we substitute the controller $u(t)=Kx(t)$ into the linearized model of Eq.(\ref{linearized}) and then the resulting closed-loop dynamical system can be rewritten in the form of a second order linear system as
\begin{eqnarray}\label{pd_design}
\ddot \sigma-K_2\dot \sigma-K_1\sigma=0,
\end{eqnarray}
where $K=[4JK_1~JK_2]$ and $K_1,K_2$ are positive constant control gain matrices. The control gain matrix $K$ is chosen as
\begin{eqnarray}\label{k_PD}
K=\left[ {\begin{array}{*{8}c}
   {-70.11} & 0 & 0 & {{-\rm{163}}{\rm{.08}}} & 0 & 0  \\
   0 & {-70.11} & 0 & 0 &  {{-\rm{163}}{\rm{.08}}} & 0  \\
   0 & 0 & {-70.11} & 0 & 0 &  {{-\rm{163}}{\rm{.08}}}  \\
\end{array}} \right],
\end{eqnarray}
\begin{figure}
\centering
\includegraphics[width=0.5\textwidth]{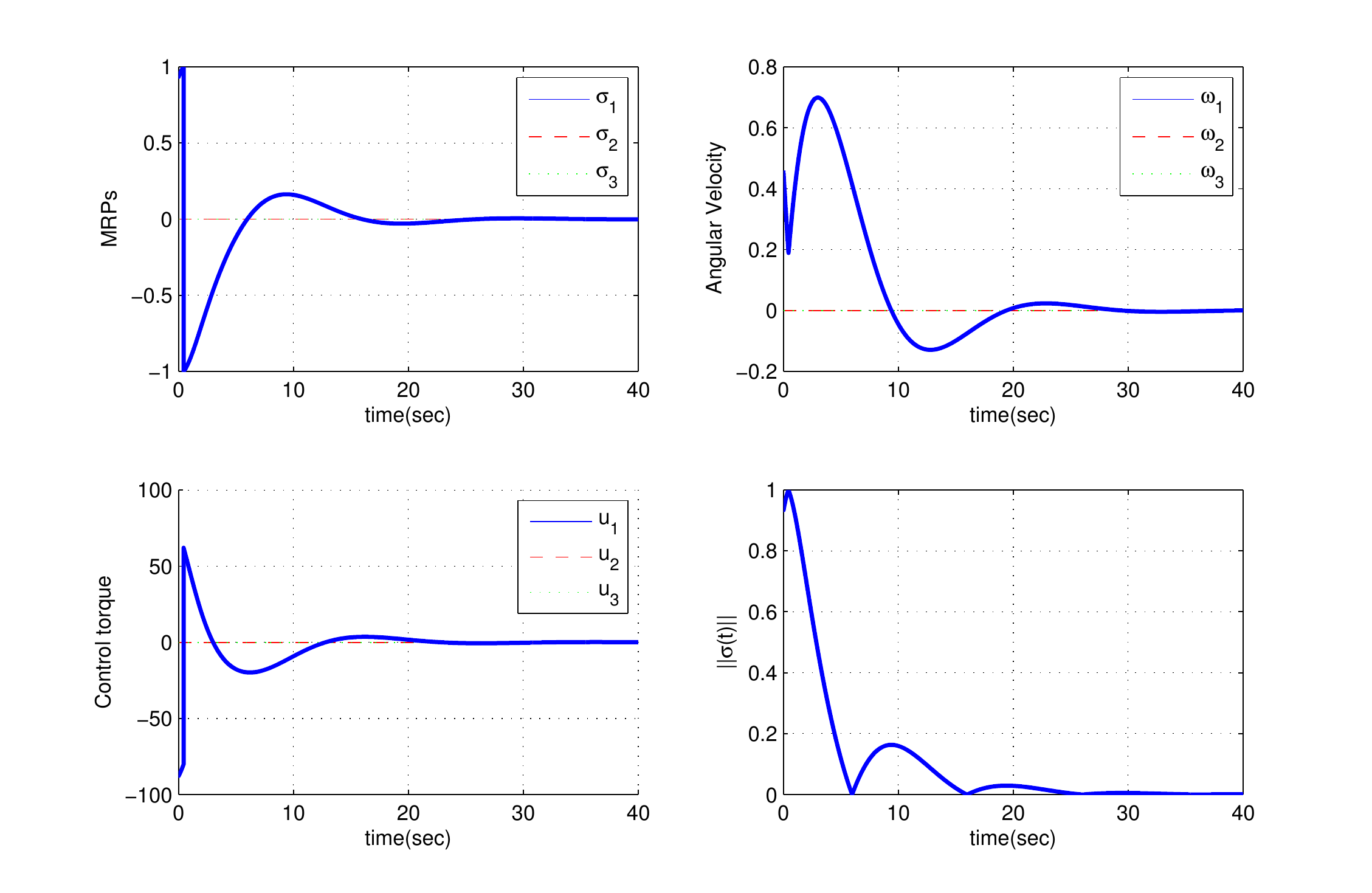}
\caption{Response of the controlled system of Eq.(\ref{nonc}) when the spacecraft is tumbling through the MRP switching point $\left\|\sigma(t)\right\|=1$ and there is no time delay in the measurement, i.e. $\tau=0$ using a simple controller $u(t)=Kx(t)$.}
\label{fig1}
\end{figure}
in order to obtain an overdamped response for the controlled system. The state variables are measured at the frequency of $1kHz$ and the results are shown Fig. (\ref{fig1}). As shown in this figure, when the spacecraft encounters a tumbling motion through $\Phi=180^\circ$ at $t=0.449(sec)$. Then, the system switches to the MRP shadow set and the control law can successfully regulate the state variables at the origin. 
Now, let us assume that there is a time delay in the measurement for the same spacecraft. Thus, we can implement the above example to the time-delayed system in Eq.(\ref{nonc}) with $\tau=0.5(sec)$ and the same initial condition and the control gain matrix $K$ as in Eq.(\ref{k_PD}). Figure (\ref{fig2}) shows the effect of this time
delay on the behavior of the controlled system response. As seen in this figure, at the beginning of the stabilization process the existence of the time delay causes the chattering phenomenon to occur for the state variables which didn't arise for the system without time delay. Moreover, it should be noted that in this simulation we are assuming that whenever $\left\|\sigma(t)\right\|=1$ the system switches to the shadow MRP set. This scenario is not a practical strategy since the presence of the time delay in the measurement causes the current value of $\sigma(t)$ is not available for the purpose of the switching to the shadow MRP set. The other strategy which is more practical is that  the switching point sets at $\left\|\sigma(t-\tau)\right\|=1$ which is the latest available information from the measurement. Then the system switches from $\sigma(t)$ to the corresponding $\sigma^s(t)$. We set $\tau=0.5(sec)$. Figure (\ref{second_stra}) shows the results of implementing this strategy. A zoomed plot of the attitude coordinates in Fig. (\ref{second_stra}) is also depicted in Fig(\ref{attitude_chatter}).  As seen in these figure,  the attitude coordinate $\sigma_1(t)$ jumps between the standard and shadow MRP sets as $t \rightarrow \infty$ and never goes to zero. It should be noted that in this set of simulations the original controlled MRP set does not go through the geometric singularity at $\Phi=360^\circ$. Thus switching to the shadow set is not required since the controller can stabilize the original MRP set to desired stable orientation, i.e, $\Phi=0^\circ$ before it reaches to the singular orientation at $\Phi=360^\circ$. This fact is shown in Fig(\ref{stable}) when we do not implement the switching rule to the system.
\begin{figure}
\centering
\includegraphics[width=0.5\textwidth]{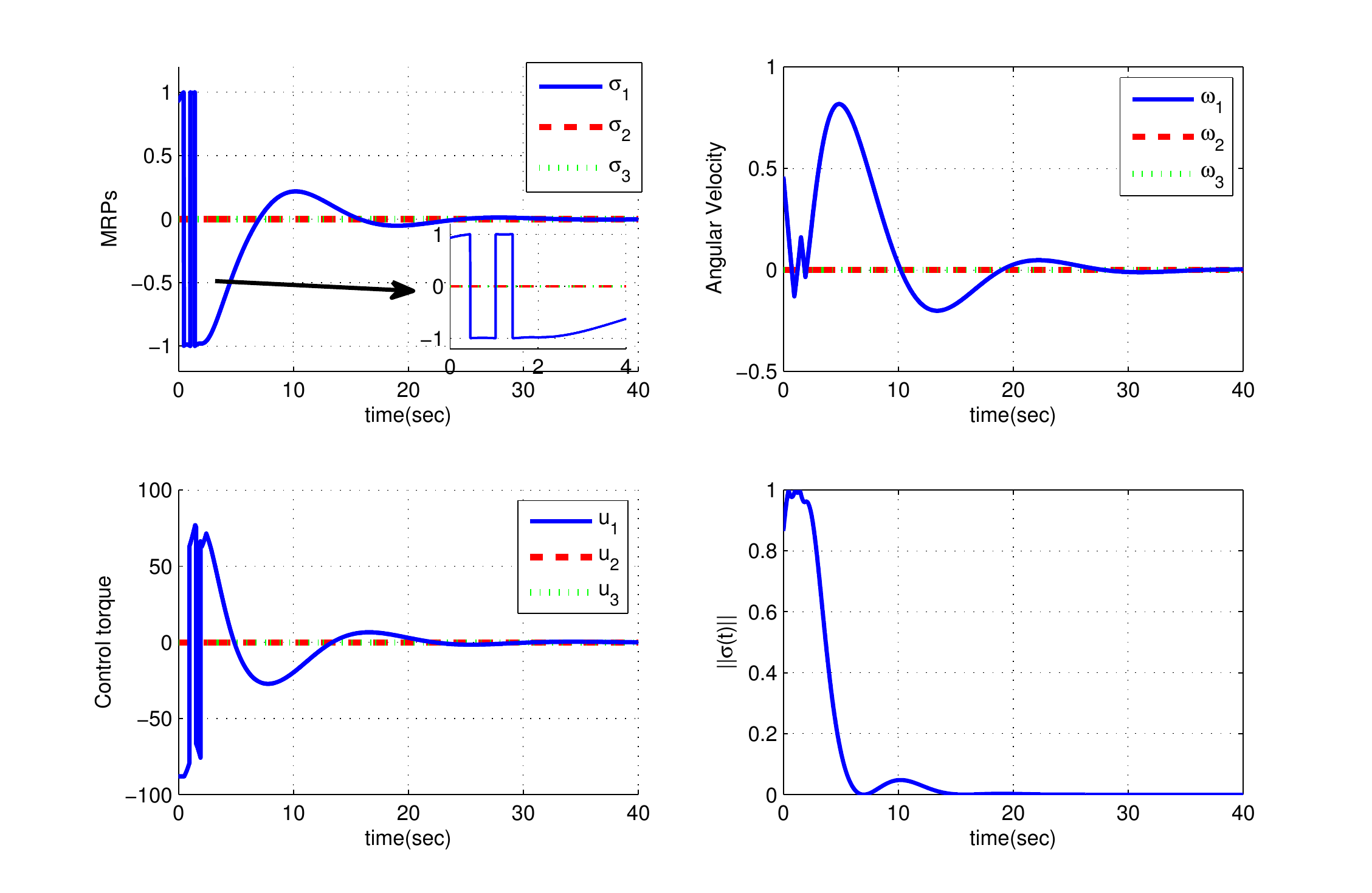}
\caption{Chattering response of the controlled system of Eq.(\ref{nonc}) when there is a time delay in the measurement and the spacecraft is tumbling through the MRP switching point $\left\|\sigma(t)\right\|=1$ using the delayed feedback controller.}
\label{fig2}
\end{figure}
\begin{figure}
\centering
\includegraphics[width=0.5\textwidth]{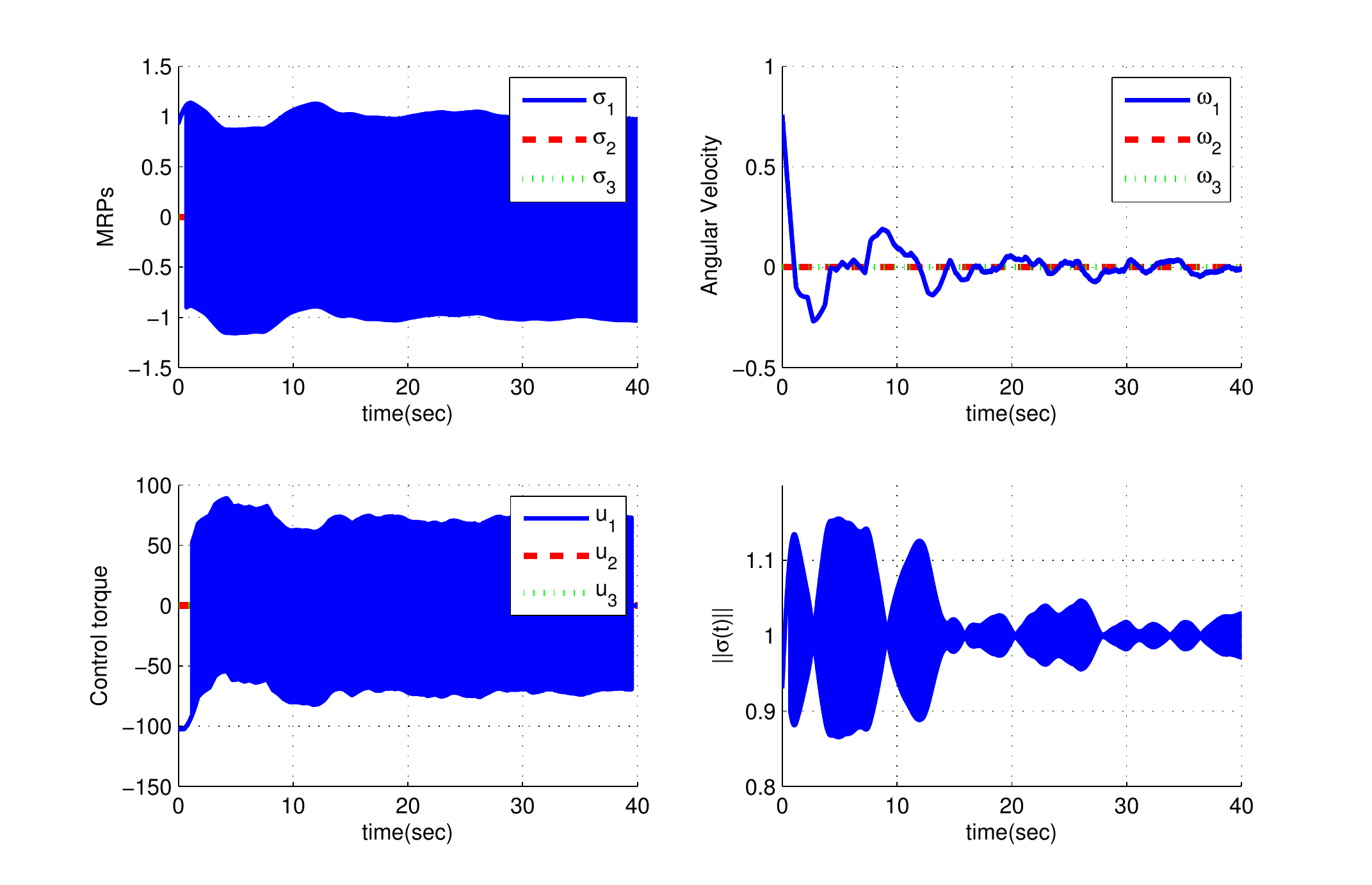}
\caption{Chattering response of the controlled system of Eq.(\ref{nonc}) when there is a time delay in the measurement and the spacecraft is tumbling through the MRP switching point $\left\|\sigma(t-\tau)\right\|=1$ using the delayed feedback controller.}
\label{second_stra}
\end{figure}
\begin{figure}
\centering
\includegraphics[width=0.4\textwidth]{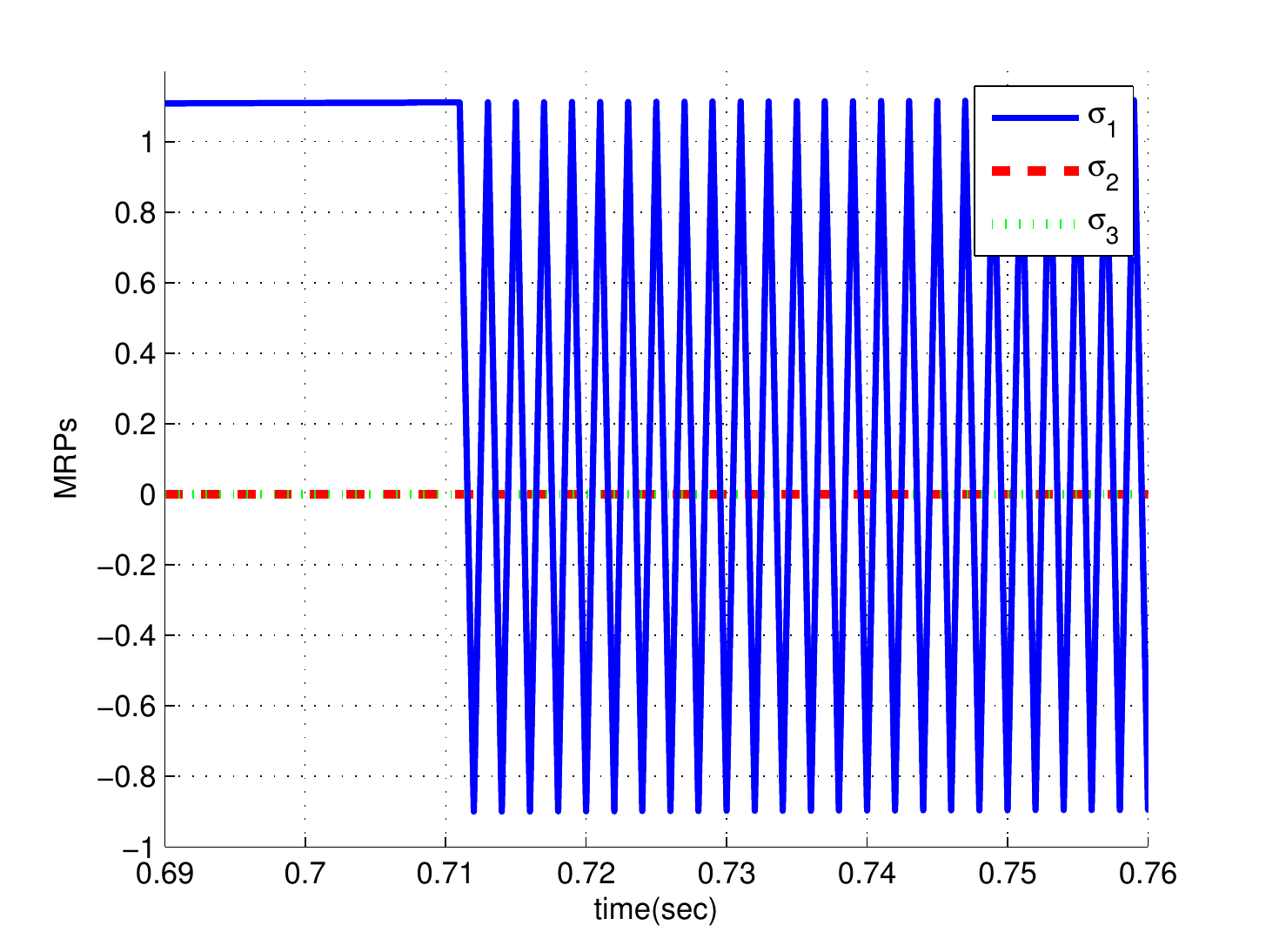}
\caption{A zoomed plot of the chattering response of the controlled attitude coordinates when there is a time delay in the measurement and the spacecraft is tumbling through the MRP switching point $\left\|\sigma(t-\tau)\right\|=1$ using the delayed feedback controller.}
\label{attitude_chatter}
\end{figure}

\begin{figure}[h!]
\centering
\includegraphics[width=0.5\textwidth]{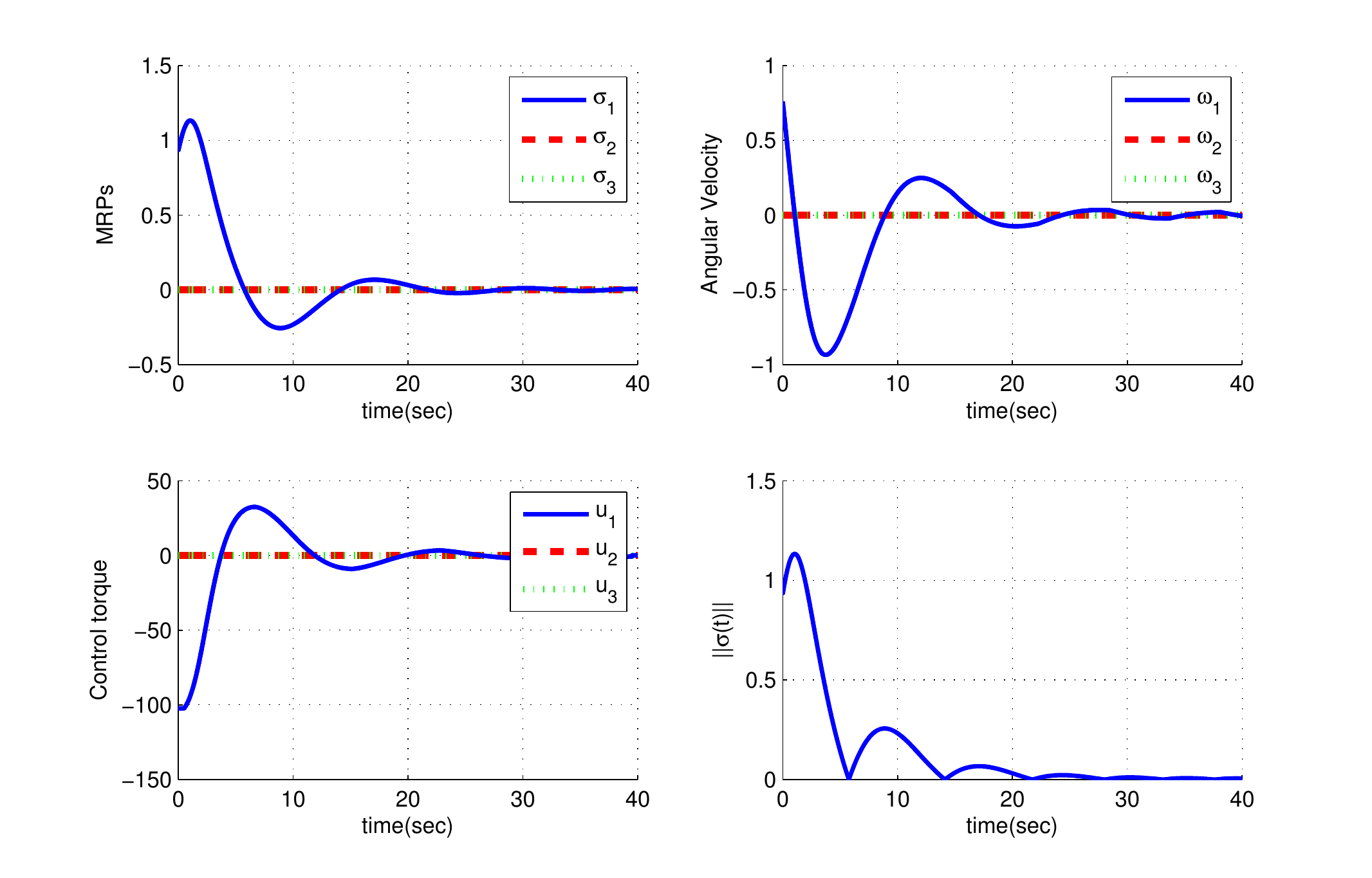}
\caption{Response of the controlled system of Eq.(\ref{nonc}) without shadow set switching using the delayed feedback controller }
\label{stable}
\end{figure}

\begin{figure}[h!]
\centering
\includegraphics[width=0.5\textwidth]{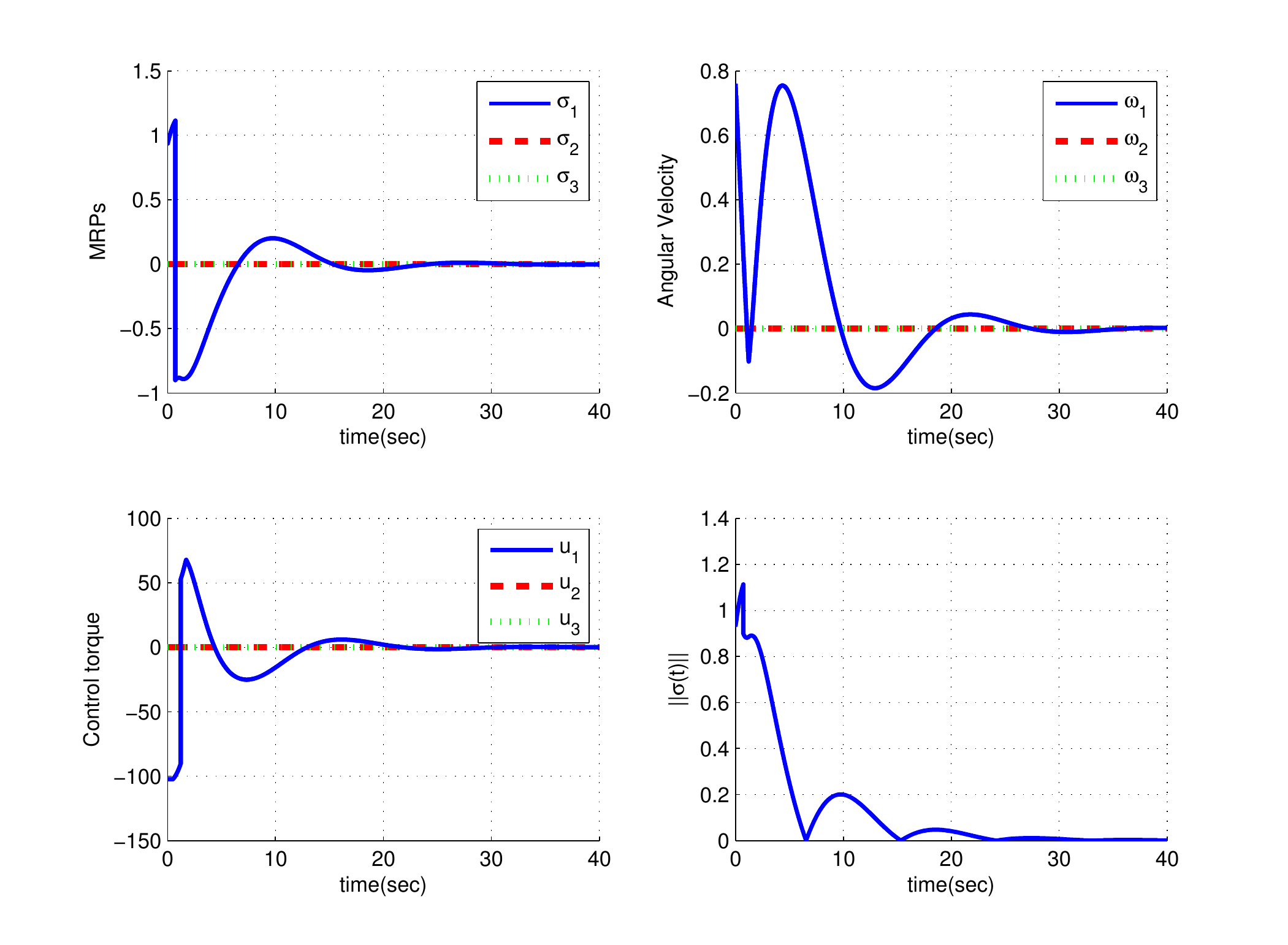}
\caption{Implementing the boundary layer switching rule to Eq.(\ref{nonc}) when there is a time delay in the measurement and the spacecraft is tumbling through the MRP switching layer $B(t)$ with $\epsilon=0.005$ using the delayed feedback controller.}
\label{BL_solution}
\end{figure}

\section{Chatter Avoidance using a Hysteretic Boundary Layer Switching Rule}
As seen in the previous section, the chattering phenomenon occurs in the response of attitude controlled system when there is a time-delay in the measurement. Different approaches are proposed for chattering avoidance in mechanical systems (see e.g., \cite{omidi2013sensitivity, omidi2015consensus, omidi2015sensitivity, omidi2014nonlinear} ). In \cite{omidi2013sensitivity}, sliding mode control (SMC) approach is used to provide the desired substrate motion trajectory for the dynamic model of nanoparticle displacement, despite the challenges in the piezoelectric substrate motion control, consisting of thermal drift, hysteresis, and other uncertainties. The objective of this section, on the other hand, is implementing the boundary layer switching rule to the closed-loop dynamical system to eliminate the chattering phenomena in the state variables of the system.
For this aim, we define a thin boundary layer in the bordering of the switching point $\left\|\sigma(t)\right\|=1$ as
\begin{eqnarray}
B(t)=\{\sigma(t) \mid 1\le \left\|\sigma(t)\right\| \le 1+\epsilon, ~ \epsilon > 0 \},
\end{eqnarray}
where $\epsilon$ is the positive constant scalar which represents the boundary layer thickness. Thus, instead of having a switching point we consider a switching layer for the MRP set. For the scenario in which switching occurs at the delayed value of the MRP set, switching to the shadow set is implemented when $\left\|\sigma(t-\tau)\right\|$ is inside the layer while on the outside of $B(t)$, the switching rule is off. Figure (\ref{BL_solution}) shows the result of implementing the boundary layer switching rule to closed-loop dynamical system of Eq.(\ref{nonc}) with the same system parameters and control gain matrix as the previous examples when $\epsilon=0.005$. As seen in this figure the chattering phenomena caused by unwanted switchings between the original and shadow MRP sets due to the existence of the time delay in the measurement is eliminated by implementing the boundary layer switching rule. It should be noted that the choice of $\epsilon$ can determine the number of unwanted chatterings in the response of the MRP set. For the large value of $\epsilon$ we may observe that the system response chatters at the beginning of stabilization process and then the controller can stabilize the closed-loop system while for small value of $\epsilon$ the chattering phenomena can be eliminated from the controlled response of the MRP set.

\section*{Acknowledgments}
Financial support from the National Science Foundation under Grant No. CMMI-1131646 is gratefully acknowledged.

\bibliographystyle{IEEEtran}
\bibliography{references}   

\begin{thebibliography}{10}
\providecommand{\url}[1]{#1}
\csname url@samestyle\endcsname
\providecommand{\newblock}{\relax}
\providecommand{\bibinfo}[2]{#2}
\providecommand{\BIBentrySTDinterwordspacing}{\spaceskip=0pt\relax}
\providecommand{\BIBentryALTinterwordstretchfactor}{4}
\providecommand{\BIBentryALTinterwordspacing}{\spaceskip=\fontdimen2\font plus
\BIBentryALTinterwordstretchfactor\fontdimen3\font minus
  \fontdimen4\font\relax}
\providecommand{\BIBforeignlanguage}[2]{{%
\expandafter\ifx\csname l@#1\endcsname\relax
\typeout{** WARNING: IEEEtran.bst: No hyphenation pattern has been}%
\typeout{** loaded for the language `#1'. Using the pattern for}%
\typeout{** the default language instead.}%
\else
\language=\csname l@#1\endcsname
\fi
#2}}
\providecommand{\BIBdecl}{\relax}
\BIBdecl

\bibitem{Marandi}
S.~R. Marandi and V.~J. Modi, ``A preferred coordinate system and the
  associated orientation representation in attitude dynamics,'' \emph{Acta
  Astronautica}, vol.~15, no.~11, pp. 833--843, 1987.

\bibitem{Shuster}
M.~D. Shuster, ``A survey of attitude representations,'' \emph{Journal of the
  Astronautical Sciences}, vol.~41, no.~4, pp. 439--517, 1993.

\bibitem{Schaub}
H.~Schaub and J.~L. Junkins, \emph{Analytical Mechanics of Space
  Systems}.\hskip 1em plus 0.5em minus 0.4em\relax AIAA, 2009.

\bibitem{Schaub3}
------, ``Stereographic orientation parameters for attitude dynamics: A
  generalization of the rodrigues parameters,'' \emph{Journal of the
  Astronautical Sciences}, vol.~44, no.~1, pp. 1--19, 1996.

\bibitem{Schaub1}
H.~Schaub, M.~R. Akella, and J.~L. Junkins, ``Adaptive control of nonlinear
  attitude motions realizing linear closed loop dynamics,'' \emph{Journal of
  Guidance, Control, and Dynamics}, vol.~24, no.~1, pp. 95--100, 2001.

\bibitem{Sharma04}
R.~Sharma and A.~Tewari, ``Optimal {N}onlinear {T}racking of {S}pacecraft
  {A}ttitude {M}aneuvers,'' \emph{IEEE Transactions of Control Systems
  Technology}, vol.~12, no.~5, pp. 677--682, 2004.

\bibitem{Sanyal1}
A.~K. Sanyal and M.~Chyba, ``Robust feedback tracking of autonomous underwater
  vehicles with disturbance rejection,'' in \emph{AAC'09, American Control
  Conference}, Honolulu, HI, June 2009.

\bibitem{karpinska}
J.~Karpi{\'n}ska and K.~Tcho{\'n}, ``Optimal extended jacobian inverse
  kinematics algorithm with application to attitude control of robotic
  manipulators,'' \emph{Robot Motion and Control 2011}, pp. 237--246, 2012.

\bibitem{Chunodkar}
A.~Chunodkar and M.~Akella, ``Attitude stabilization with unknown bounded delay
  in feedback control implementation,'' \emph{Journal of Guidance, Control, and
  Dynamics}, vol.~34, no.~2, pp. 533--542, 2011.

\bibitem{Samiei2012}
E.~Samiei, M.~Nazari, E.~A. Butcher, and H.~Schaub, ``Delayed feedback control
  of rigid body attitude using neural networks and {L}yapunov-{K}rasovskii
  functionals,'' in \emph{AAS/AIAA Spaceflight Mechanics Meeting, Charleston,
  SC, paper No. AAS 12-168}, 2012.

\bibitem{SamieiN}
E.~Samiei and E.~A. Butcher, ``Suboptimal delayed feedback attitude
  stabilization of rigid spacecraft with stochastic input torques and unknown
  time-varying delays,'' in \emph{AAS/AIAA Astrodynamics Specialist Conference,
  Hilton Head, SC, Paper No. AAS 13-837}, 2013.

\bibitem{Ailon}
A.~Ailon, R.~Segev, and S.~Arogeti, ``A simple velocity-free controller for
  attitude regulation of a spacecraft with delayed feedback,'' \emph{IEEE
  Transactions on Automatic Control}, vol.~49, no.~1, pp. 125--130, 2004.

\bibitem{Kim}
J.~Kim and J.~Crassidis, ``Robust spacecraft attitude control using model-error
  control synthesis,'' in \emph{AIAA Guidance, Navigation, and Control
  Conference, Monterey, CA}, 2002.

\bibitem{Morad1}
M.~Nazari, E.~Samiei, E.~A. Butcher, and H.~Schaub, ``Attitude stabilization
  using nonlinear delayed actuator control with an inverse dynamics approach,''
  in \emph{AAS/AIAA Spaceflight Mechanics Meeting, Charleston, SC, Paper No.
  AAS 12-237}, 2012.

\bibitem{Samiei2015attitude}
E.~Samiei, E.~A. Butcher, A.~K. Sanyal, and R.~Paz, ``Attitude stabilization of
  rigid spacecraft with minimal attitude coordinates and unknown time-varying
  delay,'' \emph{Aerospace Science and Technology}, vol.~46, pp. 412--421,
  2015.

\bibitem{Morad}
M.~Nazari, E.~A. Butcher, and H.~Schaub, ``Spacecraft attitude stabilization
  using nonlinear delayed multiactuator control and inverse dynamics,''
  \emph{Journal of Guidance, Control, and Dynamics}, vol.~36, no.~5, pp.
  1440--1452, 2013.

\bibitem{SamieiIFAC}
E.~Samiei, M.~Izadi, A.~K. Sanyal, and E.~A. Butcher, ``Delayed feedback
  asymptotic stabilization of rigid body attitude motion for large rotations,''
  in \emph{12th {IFAC} Workshop on Time Delay Systems, {A}nn {A}rbor, {MI}},
  2015.

\bibitem{Keefe}
S.~A. Oa€™Keefe and H.~Schaub, ``Shadow set considerations for modified
  rodrigues parameter attitude filtering,'' Hilton Head, SC, AAS/AIAA
  Astrodynamics Specialist Conference, Hilton Head, SC, August 2013.

\bibitem{SamieiL}
E.~Samiei, S.~Torkamani, and E.~A. Butcher, ``On {L}yapunov stability of scalar
  stochastic time-delayed systems,'' \emph{International Journal of Dynamics
  and Control}, vol.~1, no.~1, pp. 64--80, 2013.

\bibitem{omidi2013sensitivity}
E.~Omidi, A.~Korayem, and M.~Korayem, ``Sensitivity analysis of nanoparticles
  pushing manipulation by afm in a robust controlled process,'' \emph{Precision
  Engineering}, vol.~37, no.~3, pp. 658--670, 2013.

\bibitem{omidi2015consensus}
E.~Omidi and S.~N. Mahmoodi, ``Consensus positive position feedback control for
  vibration attenuation of smart structures,'' \emph{Smart Materials and
  Structures}, vol.~24, no.~4, p. 045016, 2015.

\bibitem{omidi2015sensitivity}
------, ``Sensitivity analysis of the nonlinear integral positive position
  feedback and integral resonant controllers on vibration suppression of
  nonlinear oscillatory systems,'' \emph{Communications in Nonlinear Science
  and Numerical Simulation}, vol.~22, no.~1, pp. 149--166, 2015.

\bibitem{omidi2014nonlinear}
------, ``Nonlinear vibration suppression of flexible structures using
  nonlinear modified positive position feedback approach,'' \emph{Nonlinear
  Dynamics}, vol.~79, no.~2, pp. 835--849, 2014.

\end{thebibliography}
\end{document}